# DFT analysis and FDTD simulation of $CH_3NH_3PbI_{3-x}Cl_x$ mixed halide perovskite solar cells: role of halide mixing and light trapping technique

Mohaddeseh Saffari, Mohammad Ali Mohebpour, Hamid Rahimpour Soleimani, Meysam Bagheri Tagani[*]

**Computational Nanophysics Laboratory (CNL), Department of Physics, University of Guilan, PO Box 41335-1914, Rasht, Iran**

*Corresponding Author, Email: m_bagheri@guilan.ac.ir

# DFT analysis and FDTD simulation of CH$_3$NH$_3$PbI$_{3-x}$Cl$_x$ mixed halide perovskite solar cells: role of halide mixing and light trapping technique

Mohaddeseh Saffari, Mohammad Ali Mohebpour, Hamid Rahimpour Soleimani, Meysam Bagheri Tagani[*]

**Computational Nanophysics Laboratory (CNL), Department of Physics, University of Guilan, PO Box 41335-1914, Rasht, Iran**

Abstract

Since perovskite solar cells have attracted a lot of attentions over the past years, the enhancement of their optical absorption and current density are among the basic coming challenges. For this reason, first, we have studied structural and optical properties of organic-inorganic hybrid halide perovskite CH$_3$NH$_3$PbI$_3$ and the compounds doped by chlorine halogen CH$_3$NH$_3$PbI$_{3-x}$Cl$_x$ in the cubic phase by using density functional theory (DFT). Then, we investigate the light absorption efficiency and optical current density of the single-junction perovskite solar cell CH$_3$NH$_3$PbI$_{3-x}$Cl$_x$ for the values (x=0,1,2,3) by utilizing optical constants (n,k) resulted from these calculations. The results suggest that increasing the amount of chlorine in the CH$_3$NH$_3$PbI$_{3-x}$Cl$_x$ compound leads an increase in bandgap energy, as well as a decrease in lattice constants and optical properties like refractive index and extinction coefficient of the structure. Also, the results obtained by simulation express that by taking advantage of light trapping techniques of SiO$_2$, a remarkable increase of light absorption will be achieved to the magnitude of 83.13%, which is noticeable.

1- Introduction

Fossil fuels resources limitation and the difficulties caused by greenhouse gases emission have made it necessary for everybody to attend to renewable energies such as solar energy. Therefore, due to massively utilizing solar energy, photovoltaic devices with high energy conversion are needed. So far, various substances have been used to make solar cells, which enjoy different efficiency and production cost. In this respect, perovskite solar cells are known as the most principal rivals for silicon solar cells because of their unique characteristics and high efficiency (Green et al., 2014).

The most efficient perovskite solar cell reported until now has been CH$_3$NH$_3$PbI$_3$ with efficiency of 20.1% (National Renewable Energy Laboratory, Best research-cell efficiencies). This substance is a combination of organic and inorganic parts, and results have shown if organic combination is added to perovskite structure, it will influence cell efficiency and function appropriately and noticeably (National Renewable Energy Laboratory, Best research-cell efficiencies; Kojima et al., 2009; Green et al., 2014). Many theoretical and experimental studies have been done so far regarding structural, optical and electronic properties of this substance (Wang et al., 2013; Berdiyorov et al., 2016; Berdiyorov et al., 2016; Leguy et al., 2016; Mosconi et al., 2013; Grancini et al., 2014; Chen et al., 2015). The symmetry and structure of perovskite is significantly dependent upon temperature so that in low temperatures, orthorhomrbic phase appears, higher than 161.4°k, tetragonal phase; and in temperatures higher than 330.4°k a simple cubic phase will be appeared (Wang et al., 2014; Weber, 1978). The consequences of studying perovskite in these different phases suggest that in cubic phase whose structure is more symmetrical, there are better structural and electronic properties than two other phases (Berdiyorov et al., 2016).

Perovskites show interesting properties such as ferroelectricity, thermoelectricity, piezoelectricity, birefringence, and superconductivity (Leguy et al., 2015; Grinberg et al., 2013; Guo et al., 2005; Maeno et al., 1994; Ogasawara et al., 2001; Weidenkaff et al., 2008). Organic-inorganic hybrid halide semiconductors enjoy two important and unique features i.e. powerful absorbant of light and perfect transmission of charge carriers (Zhao and Zhu, 2013). In fact, this is the feature of perovskites that caused them to be a new and significant substance in the photovoltaic industry. This substance is an ambipolar compound (Leguy et al., 2015) and in the absence of transmitting layers of hole and electron, can itself be capable of transporting charge carriers. Long excitonic diffusion lenght (Zhao et al., 2014; Stranks et al., 2013) and direct bandgap in them lead to high absorption of band edge (Green et al., 2015) and finally high efficiency of solar cell.

In designing a solar cell, the third significant and vital feature, in addition to high efficiency and low cost, is the cell stability in the nature. Despite the two features which are clear in perovskite well, no approved stability has been achieved for this substance yet; and researchers are looking for the methods to increase the stability of this substance. One of the methods to do so is halide mixing or the same dope of substance with other halogens like chlorine and bromide (Jong et al., 2016; Noh et al., 2013; Atourki et al., 2016; Zhu et al., 2016; Berdiyorov et al., 2016;). The results related to doping of $CH_3NH_3PbI_3$ with different concentrations of chlorine halogens $CH_3NH_3PbI_{3-x}Cl_x$ show that the amount of doped halogen and the alignment on which the respective halogen is apical or equatorial located affect structural and physical properties significantly (Berdiyorov et al., 2016). Also, Berdiyorov et al. (2016) studied electronic properties of $CH_3NH_3PbI_{3-x}Cl_x$ and $CH_3NH_3PbI_{3-x}Br_x$ in tetragonal phase that showed value and position of the halide mixing is important and affective on electronic properties of considered perovskites and the best result is achieved when doping is little and located at equatorial sites. Many studies have been done on band structure and electronic transmission properties of these compounds, which show organic-inorganic hybrid halide perovskite has direct bandgap whose value is appropriate for photovoltaic applications and their wide band suggests the non-localized states and easy movement and transmission of produced excitons in perovskites (Wang et al., 2013).

A typical perovskite solar cell has five different layers including anti-reflective layer, transparent conductive oxides (TCO), hole and electron transport layer (HTL and ETL) and rear metal contact. To enhance the light absorption in these cells there are few effective mechanisms: using light trapping structure (Wang et al., 2016), anti-reflection coating (Tavakoli et al., 2015), plasmonic nanoparticles (Yue et al., 2016) and etc. We have used light trapping structure in order to amplified absorption and minimize reflectance. Of course, parasitic absorption of the other layers waste plenty of trapped light which is removable by improving the quality of material, but it has been ignored in this simulation.

In this research, the structural properties, the efficiency of absorption and reflection of the solar cell made of perovskites are investigated by combining density functional theory (DFT) and finite difference time domain (FDTD) method. The dielectric constant is calculated by assuming inter-band transitions while intraband contributions are disregarded. To increase the solar cell efficiency, the light trapping techniques are used. Under optimal conditions, the light absorption efficiency attains to a high value 83.13% which is very acceptable. Section 2 is devoted to the details of the simulation. Results obtained by DFT and FDTD are presented in Section 3 and some sentences are given as a summary in conclusion part.

## 2- Method of calculations

DFT calculations: investigation of structural and optical characteristics of compounds has been done based on density function theory. At first, structures were optimized via quantum Espresso (Giannozzi et al., 2009) software with k point sampling ($8\times8\times8$). OB86-VWD-DF method (Klimesm et al. 2011) is used to take into account wan der walls interactions which are important in perovskite structures. Siesta (Soler et al., 2002) package was utilized to calculate density of state and k-point sampling is ($8\times8\times8$). GGA-PBE approximation and double-zeta-double-polarized basis sets were considered in calculations in order to extend wave functions. The mesh cutoff energy in calculations is 200Ry. Calculations regarding optical properties have been done by Siesta package and k-point mesh of ($50\times50\times50$). Optical constants like refractive index and extinction coefficient have been calculated by using real and imaginary parts of dielectric constant ($\varepsilon = \varepsilon_1 + i\varepsilon_2$) as follows:

$$n(\omega) = \frac{1}{\sqrt{2}} \left( \sqrt{\varepsilon_1^2(\omega) + \sqrt{\varepsilon_1^2(\omega) + \varepsilon_2^2(\omega)}} \right) \quad (1)$$

$$k(\omega) = \frac{1}{\sqrt{2}} \left( \sqrt{\varepsilon_1^2(\omega) - \sqrt{\varepsilon_1^2(\omega) + \varepsilon_2^2(\omega)}} \right) \quad (2)$$

FDTD simulation method for perovskite solar cell: Our simulation is based on common geometry of single junction perovskite solar cells which contains of five different layers. As shown in Fig. 1, a perovskite layer with a thickness of 350nm is limited between two layers of polymer material (HTL,ETL) which has a thickness of 15nm and 10nm, respectively. At the bottom, there is a 100nm layer of silver as a metallic electrode and as a light-reflector into the active area of the solar cell. Also in this design, a 80nm ITO (Indium tin oxide) layer and an Anti-Reflection coating (ARc) are used at the top of the cell that we optimize the thickness of the ARc to obtain the maximum efficiency. All the calculations are determined by performing the finite difference time domain (FDTD) method via common commercial proprietary softwares. In fact, using a two-dimensional simulation and periodic boundary condition in the X direction and Perfectly Matched Layer (PML) boundary condition in the Y direction, the optical reflectance and light absorptance of this perovskite solar cell are measured. In addition, a plane-polarized wave source with a wavelength range of 300-800 nanometers is used along the Y-axis to inject laterally-uniform electromagnetic energy. In this paper, PEDOT:PSS and PC60BM are assumed as HTL and ETL for hot-carriers extraction in the active layer (Wang et al., 2016). It should be mentioned that the optical constants (n, k) of the polymers, are based on measurements made via spectroscopic ellipsometry on the Ref (Lin et al., 2015).

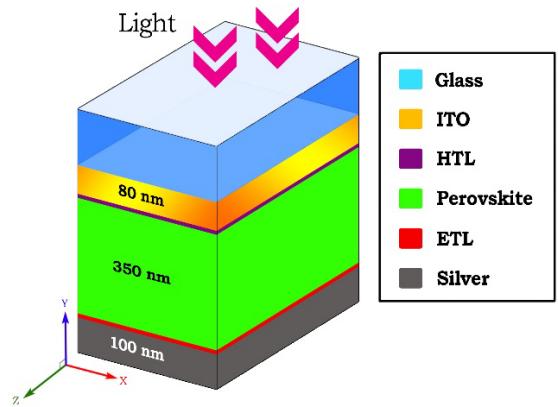

Fig. 1. three-dimensional schematic of perovskite solar cell model.

In this work, the light absorption efficiency ($A_{abs}$) is determined just like other literatures from data on $R(\lambda)$ and $T(\lambda)$ using the following equation. $R(\lambda)$ is the light reflectance from the front surface of the cell and $T(\lambda)$ is the light transmitance from the rear surface of the cell so that both quantities are calculated applying the FDTD method.

$$A_{abs}(\lambda) = (1 - R(\lambda) - T(\lambda)) \times 100\% \tag{3}$$

The short-circuit current density ($J_{SC}$) is computed as the electron charge multiplied by integrating the absorption under an AM1.5 illumination.

$$J_{sc} = \frac{e}{hc} \int \lambda \, A_{abs}(\lambda) \, I_{AM1.5}(\lambda) \, d\lambda \tag{4}$$

where e is electron charge, h is the planck's constant, and c is the speed of light in a vacuum while $\lambda$ and I are the light wavelength and solar irradiance, respectively.

### 3-1 DFT results

In the Fig. 2, band structure of four perovskite structures doped by one, two and three atoms of chlorine has been shown in symmetrical points of Brillouin zone. As seen in the figure, these perovskites have direct bandgap in cubic phase at the R point. The results show that increasing the amount of chlorine in the compound causes the bandgap to increase which has been denoted in the band structure Fig. 2. In fact, putting smaller halogen of chlorine in bigger halogen of iodine gives rise to increase the interaction between halogen and lead atom which is, in turn, very influential in the band structure properties. The interaction between p-orbital of chlorine atom with s-orbital of lead atom is greater than the interaction between p-orbital of iodine atom with s-orbital of lead atom. This lowers the maximum of valance band and increases bandgap energy. Similar behavior was observed in bromide halide mixing (Jong et al., 2016). In $CH_3NH_3PbCl_3$, the minimum of conduction band in relation to other compounds enjoys a considerable upper shift that makes more difference in bandgap of this compound than other ones. We can see a vast area both in valance band and in conduction band which represents the existence of non-localized states in the band structure (Wang et al., 2013) but this vastness in the conduction band $CH_3NH_3PbCl_3$ is less than other structures. Lattice constants decrease with increasing the amount of chlorine, which is because of putting smaller atom of chlorine with radius of 1.81 A into bigger atom of iodine with ion radius of 2.2 Å (Jong et al., 2016). After optimizing the structures, it has been seen that despite cubic phase of structures, their lattice constants are a little different in different alignments; and in fact they are quasi-cubic. As it can be seen in Table 1, in the alignments where chlorine is located, decrease of lattice constants is apparent. In $CH_3NH_3PbI_2Cl$, chlorine atom lies in the alignment of y-direction and in $CH_3NH_3PbICl_2$ the chlorines lie in the alignment of y, snd z-directions. Lattice constants and calculated bandgaps have a good agreement with previous theoretical works (Berdiyorov et al., 2016; Leguy et al., 2016) and experimental results (Leguy et al., 2016; Chen et al., 2015; Green et al., 2015).

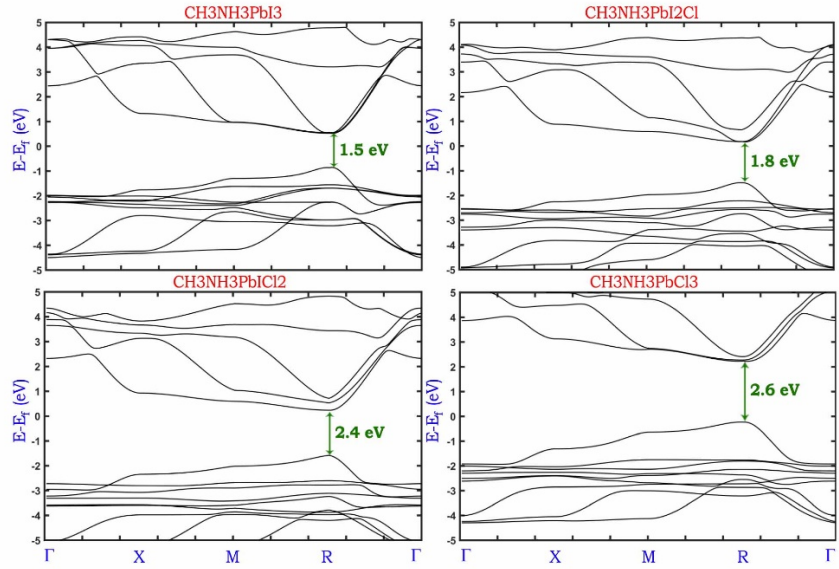

Fig. 2. Band structure of perovskite compounds.

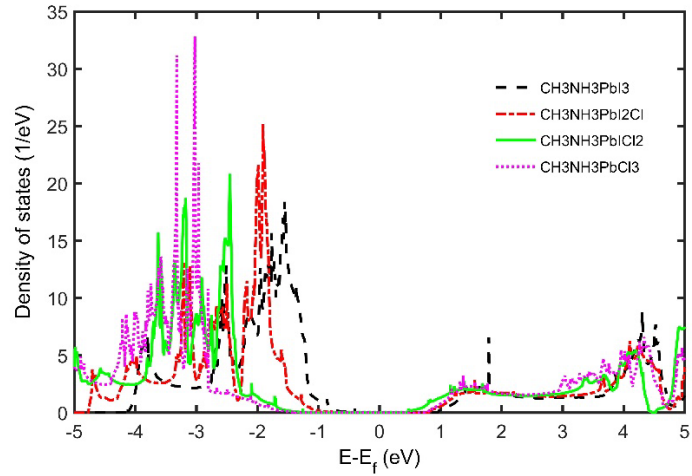

Fig. 3. Density of states of perovskites.

In order to study the structures more carefully, their density of states (DOS) have been shown in Fig. 3. On the diagram of DOS, bandgap of each structure is clear too. The area where DOS is zero represents the bandgap which is the same as designated values in band structure of each compound. The more combination between atoms' orbital, the less energy valance band locus shifts (Wang et al., 2013). As seen in the figure of density of states, the more chlorine increases, the more interaction between chlorine and lead occurs, the less energy valance band tends and bandgap of structure increases. In comparison of density of states with band structures it is observed that in energies in which the number of bands in each structure is greater and more compressed, there will be more density of states and vice versa.

Table 1: Lattice constants of considered perovskites.

| Structure | a (Å) | b (Å) | c (Å) |
|---|---|---|---|
| $CH_3NH_3PbI_3$ | 6.31 | 6.31 | 6.32 |
| $CH_3NH_3PbI_2Cl$ | 6.36 | 5.68 | 6.36 |
| $CH_3NH_3PbICl_2$ | 6.34 | 5.76 | 5.75 |
| $CH_3NH_3PbCl_3$ | 5.68 | 5.61 | 5.73 |

For investigation of optical properties of respective structures and using them in designing solar cell, optical constants i.e. refractive index and extinction coefficients have been drawn in Fig. 4(a), and (b).

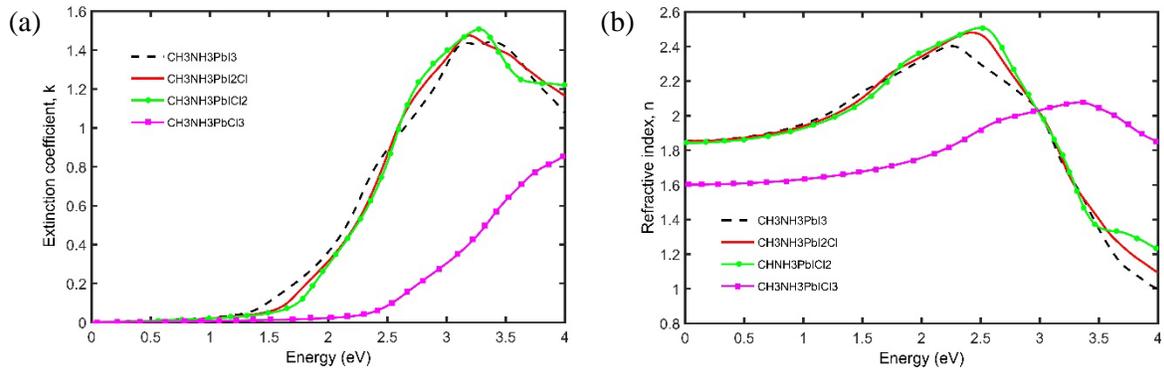

Fig. 4. Optical constants, (a) extinction coefficient. (b) refractive index.

The obtained result suggests that $CH_3NH_3PbI_2Cl$ and $CH_3NH_3PbICl_2$ have the same optical properties as $CH_3NH_3PbI_3$. The mere difference is more stability of mixing halide compounds than pure $CH_3NH_3PbI_3$. But, as seen in the diagrams, full halide $CH_3NH_3PbCl_3$ does not show desirable optical properties. It is obvious from diagrams related to the extinction coefficient, that the absorption begins to rise by increasing the energy of the bandgap. A slight absorption is seen in energies less than gap, which is caused by calculating error and is negligible. All peaks seen within the diagrams are related to passages taking place from maximum of valance band to minimum of conduction band. As it is clear in the band structure of compounds, absorption of band edge is found in energies close to gap energy which occurs from first maximum valance band to first minimum conduction band and is distinct in the diagrams related to extinction coefficient. The peak found in about 3eV energy is pertinent to the transition from first valance band to first conduction band in the X point. As seen in extinction coefficient diagram, this peak is not found in $CH_3NH_3PbCl_3$ distinctively. Extinction coefficient peaks related to this structure take place in 4.6 and 5.6 eV energies (Leguy et al., 2016), which are not within the range of energy considered by us. The peaks found in experimental works at energy about 2.5 eV (Berdiyorov et al., 2016; Leguy et al., 2016) related to transition between bands in band structure of the M point is pertinent to the share of excitons which we have not regard.

## 3-2 Numerical simulation results

As already mentioned, in this work, $CH3NH3PbI_{3-x}Cl_x$ is used as a substitute of the $CH_3NH_3PbI_3$ because appending Cl into the perovskite crystall causes more stability of the substance. In Fig. 5, the curves of light absorption efficiency versus wavelength are illustrated for four different perovskites.

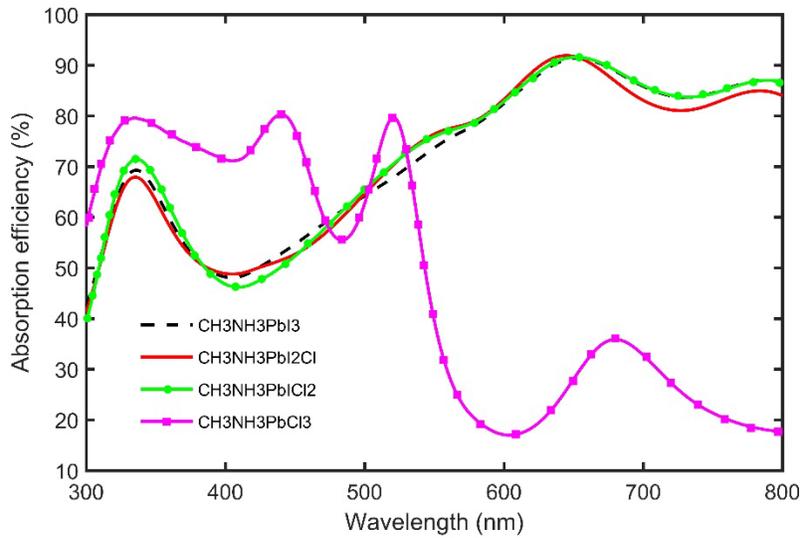

Fig. 5. The light absorption efficiency as a function of wavelength for each perovskite.

Referring to this figure we find that although the light absorption of $CH_3NH_3PbCl_3$ perovskite is much less than the $CH_3NH_3PbI_3$ but the absorptance of two other structures are as similar as triiodide structure. This property converts them to a competitive substitute for the triiodide. By averaging the amounts of light absorption of each substance, we see that the maximum absorption is belonged to $CH_3NH_3PbICl_2$ and the lowest absorption appertain to $CH_3NH_3PbCl_3$ that are equivalent to 72.21% and 47.64%, respectively. In Table 2, the ideal current density ($J_{sc}$) for these four materials in the absence of the ARc are shown. It is clear that $CH_3NH_3PbICl_2$ structure has the highest ideal current density.

Table 2: Ideal current density ($J_{sc}$) for each perovskite.

| structure | $J_{sc}$ (mA/cm$^2$) |
|---|---|
| $CH_3NH_3PbI_3$ | 20.76 |
| $CH_3NH_3PbI_2Cl$ | 20.65 |
| $CH_3NH_3PbICl_2$ | 20.84 |
| $CH_3NH_3PbCl_3$ | 10.96 |

Since this structure has a relatively premier performance, hereafter, all the results are calculated only for this structure.

### 3-3  Employment of a SiO$_2$ layer in order to mimic the reflectance of the front surface

In our perovskite solar cell, a portion of the incident light close to 15% wastes in the wavelength range 300-800nm due to reflection of the front surface of the cell. Therefore, applying of an Anti-Reflective layer can be absolutely beneficial to reduce the light losses. As shown in Fig. 1, first, we have used a simple layer of SiO$_2$ (n = 1.45) for this purpose. To achieve our aim about the most utilization of the light, we optimize the ARc thickness in Fig. 6(a). It is evident from the curves, although there is little difference in the amount of the reflected light at wavelengths larger than 600nm, we observe a reduction in the light reflection in the range of 300-600nm. By comparing these curves, we find that the best result occurs with a thickness of 60nm. In this case, the amount of short-circuit current density will increase from 20.84 to 22.74 mA/cm$^2$ just for active layer. In the same figure, black dotted line displays the light reflectance from the front surface of the solar cell in the absence of any ARc and it is quite clear that this value is too high in the wavelength range 400-500nm while has a significant decrease in the presence of an ARc. As a result of this reduction in reflection, the optical absorption of the light increases. As can be seen in Fig. 6(b), the light absorption curves as a function of wavelength have drawn either in the presence or absence of an ARc which has a thickness of 60nm. The light absorption performance at wavelengths below 600nm is improved as well just like the light reflection.

Beside reducing the light reflection, it is also useful to increase the optical path length of the incident light through the deviation of the light beam propagation direction that we have done this so well using the light trapping pyramid structures. As can be find in Fig. 7, a two-tier pyramid structure with a period of 200nm is used for light scattering. The first layer -at the bottom- is made of SiO$_2$ (n = 1.45) and the second layer of Si$_3$N$_4$ (n = 2) which have a thickness of 70 and 10nm, respectively. Also, the height of the jag is 100nm.

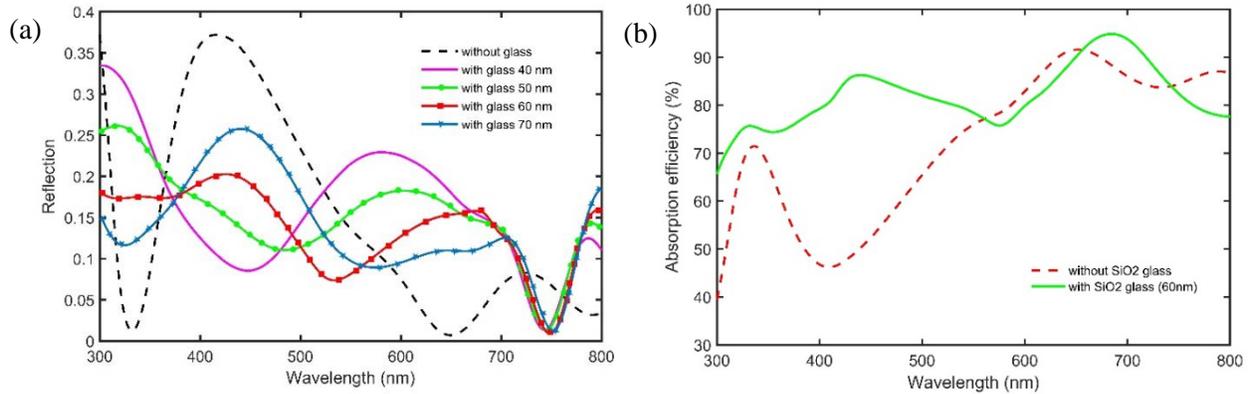

Fig. 6. (a) Wavelength-dependent light reflectance at the front surface of perovskite solar cell for four different thickness of the Anti-Reflective film. (b) Wavelength-dependent light absorption efficiency with and without simple $SiO_2$ glass.

All these parameters have been optimized and it is completely obvious that we will achieve impressive performance by smart choosing of parameters. It should be noted that according to the two-dimensional FDTD method accuracy and the geometry of the solar cells structure, we found that the improved two-dimensional FDTD method is in excellent agreement with three-dimensional simulation results. Hence, considering the high speed 2D method comparing with 3D, we still have used this way to survey the results.

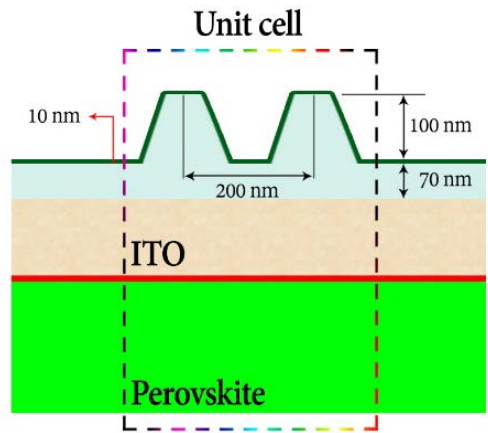

Fig. 7. two-dimensional model of the solar cell in the presence of the pyramid light trapping structure which has made from two different layers: $SiO_2$ and $Si_3N_4$.

From Fig. 8(a), we can see that the light reflection has significantly reduced from 14.85% to 10.87% by using this light trapping structure. In addition, only in the presence of this combination the $J_{sc}$ reaches its high value of 23.34 mA/cm$^2$. This scheme leads to a diminution in reflectance of the light by directing more light into active area of solar cell within the range of 400 to 600nm.

As previously mentioned, this cell escort high reflection in a color range of violet light (around 400nm). In fact, approximately 35% of the incident light at this wavelength is reflected that needs a solution. To find the best efficient thickness of $SiO_2$ in order to detract the reflection in this wavelength, the refractive index of this layer and its thickness (denote d) should satisfy the quarter wavelength formula in below. By doing this, we realize that our results are quite consistent with the theory.

$$d = \frac{\lambda}{4\eta} \tag{5}$$

Furthermore, the light absorption efficiency curves have plotted in Fig. 8(b) with and without our light trapping structure. It is clear that within the range from 300 to 600nm the optical absorption is improved. More importantly, the average of absorption efficiency in this state attains to a high value 83.13% which is very acceptable. Under these circumstances, the mixed halide perovskite $CH_3NH_3PbI_{3-x}Cl_x$ for the values (x=1,2) can be named as a main competitor for the triiodide perovskite in solar cells.

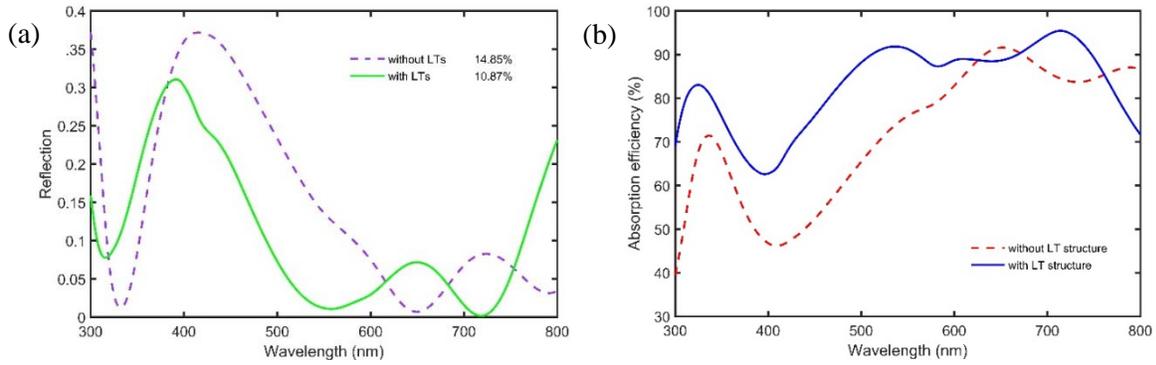

Fig. 8. (a) Wavelength-dependent light reflectance at the front surface of perovskite solar cell with and without light trapping structure. (b) Wavelength-dependent light absorption efficiency with and without light trapping structure.

**Conclusion**

Due to the importance of the perovskite structures in solar cells, structural and optical properties of organic-inorganic hybrid halide perovskite ($CH_3NH_3PbI_3$) and the compounds doped by chlorine halogen $CH_3NH_3PbI_{3-x}Cl_x$ in the cubic phase have been studied using density functional theory. In the following, by applying the material's optical constants (n, k) derived from these calculations, the light reflectance, absorptance and short-circuit current density of the single junction perovskite solar cell have been investigated using FDTD method. The results show that adding more chlorine into the $CH_3NH_3PbI_{3-x}Cl_x$ compound causes an increase in bandgap, as well as a decrease in lattice constants, refractive index and extinction coefficient of the structure. In addition, using light trapping techniques can make a significant enhancement in light absorption in these cells, up to 83.13%.